\documentclass[journal,10pt,twocolumn]{IEEEtran}
\usepackage{graphicx, times, amsmath, amsfonts, comment,amsthm}
\usepackage{amssymb, enumitem, soul, color, cite}
\usepackage[noend]{algorithmic}
\usepackage{algorithm, array}

\newcommand{\beq}{\begin{equation}}
\newcommand{\eeq}{\end{equation}}
\newcommand{\tbf}{\textbf}
\newcommand{\tit}{\textit}

\newcommand{\ud}{\mathrm{d}}

\theoremstyle{plain}

\theoremstyle{plain}

\theoremstyle{plain}

\theoremstyle{plain}

\theoremstyle{plain}

\theoremstyle{plain}

\newcommand {\Ebb}{\mathbb{E}}

\newcommand {\Pbb}{\mathbb{P}}

\newcommand {\Ccal}{\mathcal{C}}
\newcommand {\Ical}{\mathcal{I}}

\begin{document}
\title{Approximation of Meta Distribution and Its Moments for Poisson Cellular Networks
\thanks{The authors are with the Department of Electrical and Computer Engineering, University of Manitoba, Canada. 
The work was supported by NSERC.
}
}
\author{Sudarshan Guruacharya and Ekram Hossain}
\maketitle

\begin{abstract}
The notion of meta distribution as the distribution of the conditional coverage probability (CCP) was introduced in \cite{Haenggi2015}. In this letter, we show how we can reconstruct the entire meta distribution only from its moments using Fourier-Jacobi expansion. As an example, we specifically consider Poisson cellular networks. We also provide a simple closed-form approximation for its moments, along with its error analysis. Lastly, we apply the approximation to obtain a power scaling law for downlink Poisson cellular networks.
\end{abstract}


\section{Introduction}
In \cite{Haenggi2015,Haenggi2017}, the notion of meta distribution was defined for signal-to-interference ratio (SIR) coverage probability. Meta distribution essentially separates the randomness due to small-scale fading and the locations of interferers. We can trivially extend this definition to the case of signal-to-interference-plus-noise (SINR) coverage probability, which we will exclusively consider here. 

Let $\Phi$ be a point process, and let us assign to each user the conditional probability of coverage as $\Pbb(\mathrm{SINR}(\Phi) > \theta| \Phi) \equiv \Ebb[\tbf{1}(\mathrm{SINR}(\Phi)>\theta)|\Phi]$ averaged over fading for a given instantiation of the point process. Let us denote the \tit{conditional coverage probability} (CCP), which is a random variable in itself, by
\beq
\Ccal(\theta) \triangleq \Pbb(\mathrm{SINR}(\Phi) > \theta| \Phi).
\label{eqn:def-coverage}
\eeq

The \tit{meta distribution} refers to the distribution of $\Ccal$ \cite{Haenggi2015,Haenggi2017}
\beq
\bar{F}(\theta,x) = \bar{F}_{\Ccal(\theta)}(x) \triangleq \Pbb^{!0}(\Ccal(\theta)>x),
\eeq
where $x\in[0,1]$ (referred to as the {\em reliability}). We have $\bar{F}(0,x)=1$ and $\lim_{\theta \to \infty} \bar{F}(\theta,x)=0$. Also, $\bar{F}(\theta,0)=1$, and $\bar{F}(\theta,1)=0$. The $\bar{F}(\theta, x)$ is the fraction of users that achieve an SINR of $\theta$ with probability at least $x$ in each realization of $\Phi$. The  user under consideration is assumed to be located at $0$ (the origin). The mean (or standard) coverage probability is 
\beq
\Ebb^{!0}_\Phi [\Ccal(\theta)] = \Pbb^{!0}(\mathrm{SINR} > \theta) = \int_0^1 \bar{F}(\theta,x) \ud x.
\eeq
The \tit{$n$-th moment} of $\Ccal$ is the spatial average $\mu_{n}(\theta) = \Ebb^{!0}_\Phi [\Ccal^n]$.

Unfortunately, the meta distribution of $\Ccal$ is difficult to obtain in a closed analytical form. \tit{In this letter, we present a general methodology on how we can reconstruct the meta distribution from its moments using Fourier-Jacobi expansion. We examine the case of Poisson cellular network in particular, and also study a simple closed-form approximation of its moments.} The truncated Fourier-Jacobi expansion provides a better accuracy over the simple beta approximation, as used in \cite{Haenggi2015,Haenggi2017}, and its generality allows it to be used in other types of networks such as the D2D-enabled cellular and uplink cellular networks~\cite{Hesham17} as well. Likewise, the closed-form approximation of the moments allows us to make qualitative analysis of the system and rough quantitative estimates. As an example, we obtain a simple power scaling law for downlink Poisson cellular networks.


\section{Poisson Cellular Network}
Consider base stations (BSs) scattered over a two dimensional plane according to the homogeneous Poisson point process (PPP) denoted by $\Phi$ with intensity $\lambda$. Let the user under consideration, which is assumed to be at the origin, connect to the nearest BS, located $r_0$ distance away. The path-loss is assumed to be given by the power law $r_0^{-\gamma}$, where $\gamma > 2$ is the path-loss exponent. The SINR experienced by the typical user is $\mathrm{SINR} = \frac{g_0 r_0^{-\gamma} p}{\Ical + \sigma^2}$, where $\Ical = \sum_{i\in\Phi\backslash\{0\}} g_i r_i^{-\gamma} p$ is the aggregate interference from other BSs located $r_i$ distance away and transmitting over the same spectrum. Here, $p$ is the transmit power of the BSs, $\sigma^2$ is the noise power, and $g_i \sim \mathrm{Exp}(1)$ are independent and identically distributed (IID) Rayleigh fading gains. 

Given the SINR threshold $\theta$, the CCP in (\ref{eqn:def-coverage}) becomes
\begin{align}
\Ccal &= \Pbb \left( \frac{g_0 r_0^{-\gamma} p}{\Ical + \sigma^2} > \theta \bigg| \Phi \right) = \Ebb_{g_i} \left[ \Pbb \left( g_0  > \frac{\theta r_0^{\gamma}}{p} (\sigma^2 + \Ical) \bigg| g_i, \Phi \right) \right], \nonumber \\ 
&\stackrel{(a)}{=} \exp \left( -\frac{\theta \sigma^2 r_0^{\gamma}}{p} \right) \Ebb_{g_i} \left[ \exp \left( - \theta \sum_{i\in\Phi\backslash\{0\}} g_i \left(\frac{r_0}{r_i}\right)^{\gamma} \right) \right], \nonumber \\
&\stackrel{(b)}{=} \exp \left( -\frac{\theta \sigma^2 r_0^{\gamma}}{p} \right) \prod_{i\in\Phi\backslash\{0\}} \left[ 1 + \theta \left(\frac{r_0}{r_i}\right)^{\gamma} \right]^{-1}, 
\label{eqn:coverage-rayleigh}
\end{align}
where $(a)$ is due to $g_0 \sim \mathrm{Exp}(1)$, and $(b)$ is the Laplace transform of the sum of IID exponential random variables. 

To find the $n$-th moments of $\Ccal$, we take the spatial average $\mu_n = \Ebb_\Phi^{!0}[\Ccal^n]$. From (\ref{eqn:coverage-rayleigh}), conditioning on $r_0$ and following the procedure given in \cite[Theo. 2]{Haenggi2015}, we have
\begin{align*}
\mu_n = \Ebb_{r_0} \Big[ e^{-\frac{n \theta \sigma^2}{p} r_0^{\gamma}} \; e^{ -\pi\lambda \rho_n r_0^2 } \Big],
\end{align*}
where $\rho_n = 2\int_0^1 (1 - v(y))y^{-3} \ud y$ and $v(y) = (1+\theta y^\gamma)^{-n}$.
For PPP, the density of the distance to the nearest BS is given by $f_{r_0}(r) = 2 \pi r \lambda e^{-\pi \lambda r^2}$. Hence, 
\begin{align*}
\mu_n &= 2 \pi \lambda \int_0^\infty e^{-\frac{n \theta \sigma^2 }{p} r^{\gamma}} e^{- \pi \lambda \rho_n r^2 }  e^{-\pi \lambda r^2} r \ud r.
\end{align*}
Changing the variable to $z = r^2$, we have the final form of the integral as
\beq
\mu_n = \pi \lambda \int_0^\infty \exp \{-(A_n z + B_n z^{\gamma/2})\} \ud z,
\label{eqn:nth-moment}
\eeq
where $A_n = \pi \lambda (\rho_n + 1)$ and $B_n = \frac{n \theta \sigma^2}{p}$. From \cite[Theo. 2]{Haenggi2015}, we also have $1+\rho_n = {}_2F_1(n,-2/\gamma; 1-2/\gamma;-\theta)$. 



\section{Reconstructing the Meta Distribution by Fourier-Jacobi Expansion}

Since meta-distribution has a finite support of $[0,1]$, the problem of reconstructing the distribution from its moments fits into the Hausdorff moment problem and can be done using the Jacobi polynomials~\cite{Szego1975}\cite[Ch. 18]{Olver2010}. Since canonically the Jacobi polynomials, $P_n^{*(\alpha, \beta)}$, are defined over $[-1, 1]$, we will instead be working with shifted Jacobi polynomials. The shifted Jacobi polynomial, $P_n^{(\alpha, \beta)}$, is then defined over $[0,1]$ with respect to the weight function $w(x) = (1-x)^{\alpha} x^{\beta}$, where $\alpha, \beta > -1$. The two polynomials are related by the identity $P_n^{(\alpha, \beta)}(x) = P_n^{*(\alpha, \beta)}(2x-1)$ for $x\in[0,1]$. Facts about $P_n^{*(\alpha, \beta)}$ can thus be converted into facts about $P_n^{(\alpha, \beta)}$ by change of variable.

A number of orthogonal polynomials occur as special cases of the Jacobi polynomial. When $\alpha = \beta = 0$, the Jacobi polynomial reduces to Legendre polynomial; when $\alpha = \beta = \pm 1/2$, it reduces to Chebyshev polynomial; when $\alpha = \beta$, it reduces to Gegenbaur polynomial. These polynomials satisfy the orthogonality condition 
\beq
\int_0^1 P_n^{(\alpha, \beta)}(x) P_m^{(\alpha, \beta)}(x) w(x) \ud x = h_n \delta_{mn},  
\label{eqn:orthogonality-cond}
\eeq 
where $\delta_{mn}$ is Kronecker delta function and $h_n$ is the normalization constant given by (see \cite[Ch 18.3]{Olver2010} for canonical)
\[ h_n = \frac{1}{2n + \alpha + \beta + 1} \frac{\Gamma(n+\alpha+1)\Gamma(n+\beta+1)}{n! \Gamma(n+\alpha+\beta+1)}. \]

As a check, when $n=0$, we have 
\[ h_0 = \frac{1}{\alpha+\beta+1} \frac{\Gamma(\alpha+1)\Gamma(\beta+1)}{\Gamma(\alpha+\beta+1)} = \frac{\Gamma(\alpha+1)\Gamma(\beta+1)}{\Gamma(\alpha+\beta+2)}, \] 
which is the normalization constant for beta distrbution $\mathrm{Beta}(\beta+1,\alpha+1)$. Also, when $\alpha=\beta=0$, we have $h_n = \frac{1}{2n+1}$, which matches with the normalization constant for shifted Legendre polynomial.

The explicit expression for the shifted Jacobi polynomial is given by (see \cite[Eqn 18.5.8]{Olver2010} for canonical)
\beq 
P_n^{(\alpha,\beta)}(x) = \sum_{\ell=0}^n \binom{n+\alpha}{\ell} \binom{n+\beta}{n-\ell} x^\ell (x-1)^{n-\ell}. 
\label{eqn:explicit-shifted-jacobi-poly}
\eeq
In particular, $P^{(\alpha,\beta)}_0(x) = 1$. 

Given all the moments of the meta distribution $\mu_n = \int_0^1 x^n f_\Ccal(x) \ud x$, we can reconstruct the PDF of the meta distribution defined over the interval $[0,1]$ using the shifted Jacobi polynomials via Fourier-Jacobi expansion as:
\beq 
f_\Ccal(x) = w(x) \sum_{n=0}^\infty a_n P_n^{(\alpha,\beta)}(x). 
\label{eqn:fourier-jacobi-pdf}
\eeq

As with the usual Fourier expansion, we can extract the coefficients $a_n$ by multiplying both sides by $P_n^{(\alpha,\beta)}(x)$, integrating with respect to $x$, and applying the orthogonality condition (\ref{eqn:orthogonality-cond}). Using (\ref{eqn:explicit-shifted-jacobi-poly}), this gives us
\begin{align}
a_n &= \frac{1}{h_n} \int_0^1 f_\Ccal(x) P_n^{(\alpha,\beta)}(x) \ud x, \nonumber \\
&= \frac{1}{h_n} \sum_{\ell=0}^n \binom{n+\alpha}{\ell} \binom{n+\beta}{n-\ell} \hat{\mu}_{n\ell}, \label{eqn:fourier-jacobi-coeff}
\end{align}
where $\hat{\mu}_{n\ell} = \int_0^1 x^\ell (x-1)^{n-\ell} f_\Ccal(x) \ud x$ are the modified moments. The modified moments  $\hat{\mu}_{n\ell}$ are related to the usual moments $\mu_n$ by the binomial expansion
\begin{align}
\hat{\mu}_{n\ell} &= \int_0^1 x^\ell \sum_{k=0}^{n-\ell} \binom{n-\ell}{k} (-1)^k x^{n-\ell-k} f_\Ccal(x) \ud x, \nonumber \\
&= \sum_{k=0}^{n-\ell} \binom{n-\ell}{k} (-1)^k \int_0^1 x^{n-k} f_\Ccal(x) \ud x,  \nonumber \\
&= \sum_{k=0}^{n-\ell}  \binom{n-\ell}{k} (-1)^k \mu_{n-k}. \label{eqn:fourier-jacobi-mod-coeff}
\end{align}
In particular, $\hat{\mu}_{00} = 1$ and $a_0 = 1/h_0$. Hence, we have completely reconstructed the PDF of meta distribution from its moments using shifted Jacobi polynomials. The first term in the series (\ref{eqn:fourier-jacobi-pdf}) is indeed given by the beta distribution.

We can also integrate the PDF to obtain the CDF, which is more useful in practical applications. We have 
\begin{align*}
F_\Ccal(x) &= \int_0^x f_\Ccal(x) \ud x = \sum_{n=0}^\infty a_n \int_0^x w(x) P_n^{(\alpha,\beta)}(x) \ud x.
\end{align*}        

The shifted Jacobi polynomials can also be generated by the Rodrigues' formula 
\beq 
P_n^{(\alpha, \beta)}(x) = \frac{(-1)^n}{n!} \frac{1}{(1-x)^\alpha x^\beta} \frac{\ud^n}{\ud x^n} \left[ (1-x)^{\alpha+n} x^{\beta+n} \right]. 
\label{eqn:rodrigues-formula}
\eeq
Using the Rodrigues' formula (\ref{eqn:rodrigues-formula}), it can be shown that the integral for $n \geq 1$ is
\[ \int_0^x (1-x)^{\alpha} x^{\beta} P_n^{(\alpha,\beta)}(x) \ud x  = -\frac{1}{n} (1-x)^{\alpha+1} x^{\beta+1} P_{n-1}^{(\alpha+1, \beta+1)}(x). \]

Therefore, the required expansion for the CDF is 
\begin{align} 
F_\Ccal(x) &= \frac{1}{h_0} \int_0^x (1-x)^\alpha x^\beta \ud x \nonumber \\
& - \sum_{n=1}^\infty \frac{a_n}{n} (1-x)^{\alpha+1} x^{\beta+1} P_{n-1}^{(\alpha+1, \beta+1)}(x).  \label{eqn:fouier-jacobi-cdf}
\end{align}

While any value of $\alpha$ and $\beta$ can be considered, so long as they are greater than $-1$, it is prudent to take their values such that the first two correction terms vanishes by setting $a_1 = a_2 = 0$. The values of $\alpha$ and $\beta$ thus obtained corresponds to the values obtained by moment matching method for the beta distribution:
\begin{align*} 
\alpha + 1 &= \frac{(\mu_1 - \mu_2)(1 - \mu_1)}{\mu_2 - \mu_1^2}, \quad 
\beta +1 = \frac{(\alpha+1)\mu_1}{1-\mu_1}. 
\end{align*}
Thus, we have simultaneously justified the use of moment matching method, while at the same  improving on it.

The series (\ref{eqn:fourier-jacobi-pdf}) will in general not converge without imposing some side condition on $a_n$. The convergence of series can be investigated using Weierstrass M-test. Without losing any generality, let $-1 < \beta \leq \alpha$ such that $\alpha \geq -1/2$. Then, from \cite[Eqn 18.14.1]{Olver2010}, we have $|P_n^{(\alpha,\beta)}(x)| \leq P_n^{(\alpha,\beta)}(1) = \frac{(\alpha+1)_n}{n!}$, where $(.)_n$ is the rising factorial. Here, $\frac{(\alpha+1)_n}{n!} = \prod_{k=1}^n (1 + \frac{\alpha}{k})$. When $\alpha \leq 0$, we have $\prod_{k=1}^n (1 + \frac{\alpha}{k}) \leq (1 + \frac{\alpha}{n})^n \leq e^{\alpha}$. Similarly, when $\alpha > 0$, we have $\prod_{k=1}^n (1 + \frac{\alpha}{k}) \leq (1 + \alpha)^n \leq e^{\alpha n}$. We can now upper bound each term of the series as 
\[ |a_n P_n^{(\alpha,\beta)}(x)| \leq \left\{ \begin{array}{lcr} 
|a_n| e^{\alpha} & \mathrm{for} & \alpha \leq 0, \\
|a_n| e^{\alpha n} & \mathrm{for} & \alpha > 0.
\end{array} \right.
\]
The series will converge absolutely and uniformly: if $\sum_n |a_n|$ converges for $\alpha \leq 0$; or if we can express $|a_n|$ as $|a_n| = b_n e^{-\bar{\alpha} n}$ where $\bar{\alpha} \geq \alpha$ and $b_n \geq 0$ such that $\sum_n b_n$ converges for $\alpha > 0$.

\section{Approximate Moments of $\Ccal$ and Error Analysis}
\subsection{Approximate Moments of $\Ccal$}
The integral (\ref{eqn:nth-moment}) does not have a closed-form solution. Nevertheless, a simple closed-form approximation can be given as \cite[Eqn 4]{Guruacharya2016}
\beq
\mu_n \simeq \pi \lambda \left[ A_n + \frac{\gamma B_n^{2/\gamma}}{2 \Gamma(\frac{2}{\gamma})} \right]^{-1}.
\label{eqn:nth-moment-approx}
\eeq

One importance of this formula lies in the fact that we can explicitly solve for the transmit power $p$ given the coverage constraint 
\beq
\Pbb^{!0}(\Ccal > x) \geq 1 - \epsilon,
\label{eqn:QoS-constraint}
\eeq
where $\epsilon \in (0,1)$ is some arbitrary value which represents the quality-of-service. Since $|\Ccal| \leq 1$, we have the special case of lower Markov bound as 
\cite[Sec. 6.2.a]{Lin2011} $\Pbb^{!0}(\Ccal \geq x) \geq \mu_2 - x^2$,
where $\mu_2 \geq x^2$. The constraint (\ref{eqn:QoS-constraint}) is always satisfied if $\mu_2 - x^2 \geq 1 - \epsilon$. Using (\ref{eqn:nth-moment-approx}) for $\mu_2$ and after some basic algebra, the minimum $p$ is
\beq
p \simeq c \lambda^{-\gamma/2},
\eeq
where $c = \left[\frac{2\pi(1 - (1 - \epsilon + x^2) (1+\rho_2))}{\gamma(1 - \epsilon + x^2)(2 \theta \sigma^2)^{2/\gamma}} \Gamma \left(\frac{2}{\gamma}\right) \right]^{-\gamma/2}$. This gives a simple power scaling law based on the meta-distribution. Similar scaling laws were obtained in \cite{Sanguanpuak2018} using the first moment. In the following, we will conduct the error analysis of the above approximation.

\subsection{Error Analysis}
In \cite{Guruacharya2016}, we had the following: For any positive constants $A >0$, $B > 0$ and $\gamma > 2$, let 
\begin{align} 
I &= \int_0^\infty \exp\{-(Az + Bz^{\gamma/2})\} \ud z \\
\mathrm{and} \quad \hat{I} &= \frac{1}{K}, \quad \mathrm{where} \quad K = A + \frac{\gamma B^{2/\gamma}}{2 \Gamma(\frac{2}{\gamma})}, \label{eqn:def-K}
\end{align}
then we have the integral approximation $I \simeq \hat{I}$.

The approximation is exact when $A=0$ or $B=0$ or $\gamma = 2$. However,  the error of this approximation was not analyzed in \cite{Guruacharya2016}. To do so, first observe that we can equivalently express $\hat{I}$ in an integral form as 
\[ \hat{I} = \frac{1}{K} = \int_0^\infty e^{-Kz} \ud z. \] 

Now, let $h(z) = Az + B z^{\gamma/2}$ and $g(z) = K z$, so that their difference is $f(z) = h(z) - g(z) = (A-K)z + Bz^{\gamma/2}$. Thus, our required integral can be expressed as
\begin{align*}
I &= \int_0^\infty e^{-h(z)} \ud z =  \int_0^\infty e^{-f(z)} e^{-g(z)} \ud z.
\end{align*}
Integrating by parts, we have 
\begin{align*}
I &= \left. e^{-f(z)} \int e^{-g(z)} \ud z \right|_0^{\infty} - \int_0^\infty f'(z) e^{-f(z)} \left[ \int e^{-g(z)} \ud z \right] \ud z \\
&= \left. - e^{-f(z)} \frac{e^{-Kz}}{K} \right|_0^{\infty} + \frac{1}{K} \int_0^\infty f'(z) e^{-f(z)} e^{-Kz} \ud z.
\end{align*}

Now, regardless of whether $A-K$ is positive or negative, $f(\infty) = \infty$ because $\gamma>2$. Since $e^{-f(\infty)} = 0$ and $e^{f(0)} = 1$, the integral reduces to
\begin{align*} 
I &=  \frac{1}{K} + \frac{1}{K} \int_0^\infty f'(z) e^{-f(z)} e^{-Kz} \ud z 
\end{align*}

Recalling that $\hat{I} = 1/K$, the approximation error is now 
\begin{align}
|I - \hat{I}| &= \frac{1}{K} \left| \int_0^\infty f'(z) e^{-f(z)} e^{-Kz} \ud z  \right|, \nonumber \\
&\leq \frac{1}{K} \int_0^\infty \left| f'(z) e^{-f(z)} e^{-Kz} \right| \ud z. \label{eqn:initial-error-bound}
\end{align}

If the maximum value attained by $e^{-f(z)}$ is denoted by $M = \max_{z\geq0} \, e^{-f(z)} < \infty$, then we have further inequality
\begin{align}
|I - \hat{I}| &\leq \frac{M}{K} \int_0^\infty \left| f'(z) e^{-Kz} \right| \ud z \nonumber \\
&\stackrel{(a)}{\leq} \frac{M}{K} \int_0^\infty \left[ |A-K| + \frac{\gamma}{2} B z^{\gamma/2 - 1} \right] e^{-Kz} \ud z, \nonumber \\
&\stackrel{(b)}{=} \frac{M}{K} \left[ \frac{|A - K|}{K} + \frac{\gamma B}{2} \frac{\Gamma(\gamma/2)}{K^{\gamma/2}} \right], \nonumber \\
&\stackrel{(c)}{=} \frac{\gamma M}{2K} \left[ \frac{1}{\Gamma(2/\gamma)} \frac{B^{2/\gamma}}{K} +  \Gamma\left(\frac{\gamma}{2}\right) \left( \frac{B^{2/\gamma}}{K} \right)^{\gamma/2} \right], \label{eqn:final-error-bound}
\end{align}
where $(a)$ follows from triangle inequality,  
$(b)$ follows by term wise integration, and $(c)$ follows from the definition of $K$ in (\ref{eqn:def-K}). Equation (\ref{eqn:final-error-bound}) gives us our required error bound.

As the final piece of analysis, we need to evaluate the maximum $M$, which we assumed to be finite. We will now show that $M$ is indeed finite and independent of $A$ and $B$. First, observe that 
\begin{align*} 
f(z) &= (A-K)z + Bz^{\gamma/2} = - \frac{B^{2/\gamma}}{\frac{2}{\gamma} \Gamma(\frac{2}{\gamma})} z + Bz^{\gamma/2},  \\
f'(z) &=  - \frac{B^{2/\gamma}}{\frac{2}{\gamma} \Gamma(\frac{2}{\gamma})} + \frac{\gamma}{2} B z^{\gamma/2 -1}, \\
f''(z) &= \frac{\gamma}{2} \left( \frac{\gamma}{2} - 1\right) B z^{\gamma/2 - 2}.
\end{align*}
Since $B>0$ and $\gamma > 2$, we have $f''(z) \geq 0$ for all $z \geq 0$. Hence, $f(z)$ is a convex function over $z \geq 0$, with 
a unique minima at 
\[ z^* = 1 \big/ \left[ \Gamma^{\frac{2}{\gamma} - 2} \left(\frac{2}{\gamma}\right) B^{2/\gamma} \right]. \]
obtained by solving $f'(z) = 0$. Thus, the minimum value attained by $f(z)$ is, after some simplification,
\[ \min_z \, f(z) = \left( 1- \frac{\gamma}{2} \right) \Big/ \Gamma^{\frac{\gamma}{\gamma-2}} \left(\frac{2}{\gamma}\right), \]
which is independent of $A$ and $B$. Therefore, the required expression for $M$ is $M = \max_z \, e^{-f(z)} = e^{- \min_z f(z)} = \exp \{ \left( 1- \frac{\gamma}{2} \right) \big/ \Gamma^{\frac{\gamma}{\gamma-2}} \left(\frac{2}{\gamma}\right) \}$.

In the above error expression (\ref{eqn:final-error-bound}), as $A \to \infty$ while $B$ is fixed, $K \to \infty$ as well. Thus, $|I - \hat{I}| \to 0$. Likewise, as $B \to \infty$ while $A$ is fixed, we have $B^{2/\gamma}/K \to (2/\gamma) \Gamma(2/\gamma)$ because $\gamma/2 > 1$. Also, since $K \to \infty$ as $B \to \infty$, we thus have $|I - \hat{I}| \to 0$. To conclude, $|I - \hat{I}| \to 0$ either (i) when $A \to 0$ or $A \to \infty$ and $B$ is fixed, or (ii) when $B \to 0$ or $B \to \infty$ and $A$ is fixed.

Likewise, from (\ref{eqn:initial-error-bound}) we also have

\[ \left| \frac{I}{\hat{I}} - 1 \right| = \left| \int_0^\infty f'(z) e^{-f(z)} e^{-Kz} \ud z  \right|. \] 

Similar argument as before can be used to show that $I/\hat{I} \to 1$ when $A$ goes to infinity. However, when $B$ goes to infinity, the error is bounded by $O(1)$.

\section{Numerical Results}
We consider the following parameters: $\lambda = 0.001$ per $\mathrm{m}^2$, $\gamma = 5$, $\theta = 0$ dB, $p = 0$ dBm, $\sigma^2 = -100$ dBm. Points are uniform randomly dropped over an area of $\pi \times 500^2$ $\mathrm{m}^2$, with the typical user located at the origin. For every realization of the point process, 700 random channel realizations are used to find the conditional coverage probability of the typical user. Likewise, 5,000 geometric configurations are used to construct the meta distribution.

\begin{figure}[t]
	\begin{center}
		\includegraphics[width=\columnwidth] {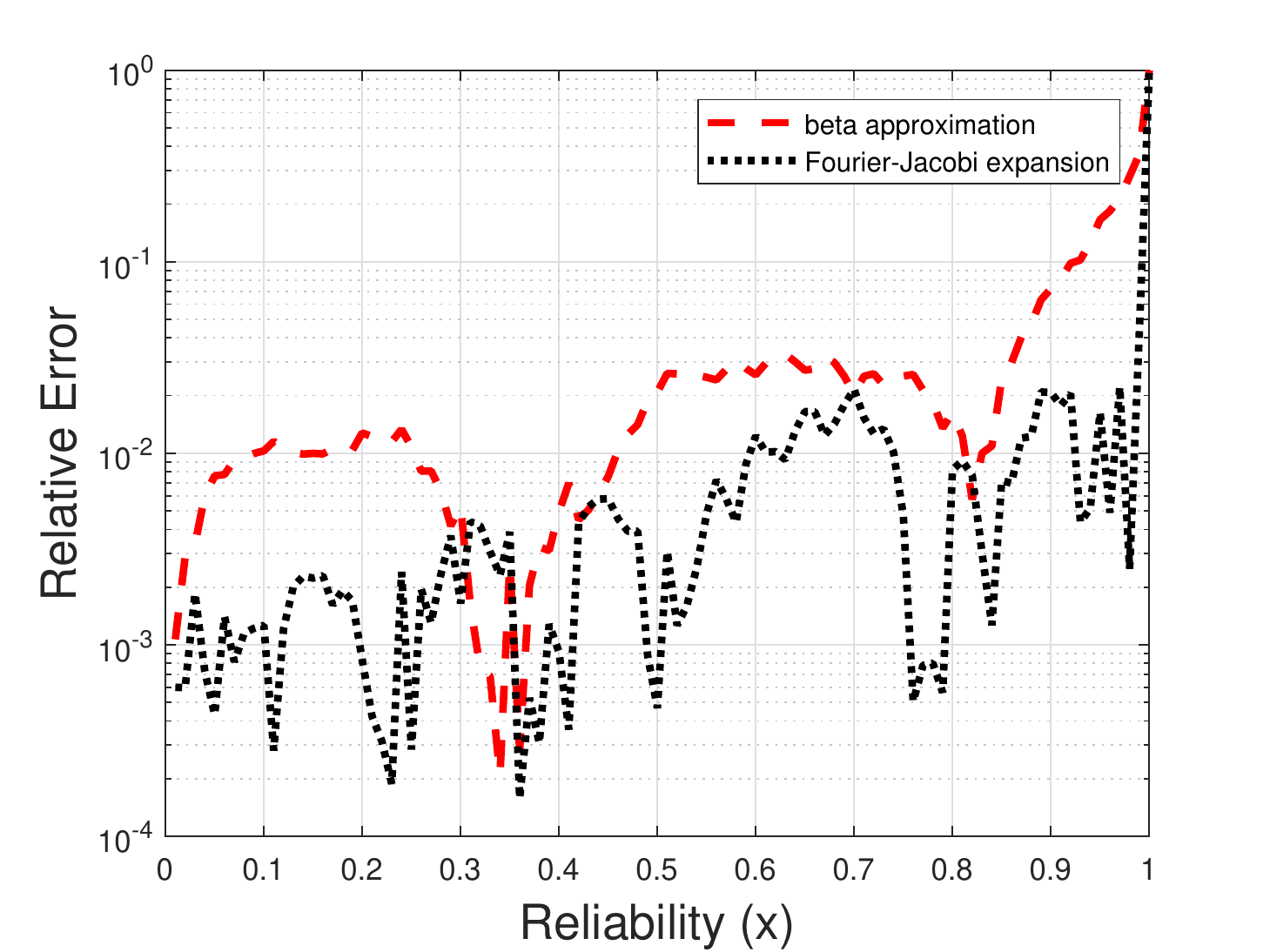}
		\caption{Relative error versus reliability for $\gamma = 5$.}
		\label{fig:relative-error}
	\end{center}
\end{figure}

\begin{figure}[t]
	\begin{center}
		\includegraphics[width=\columnwidth] {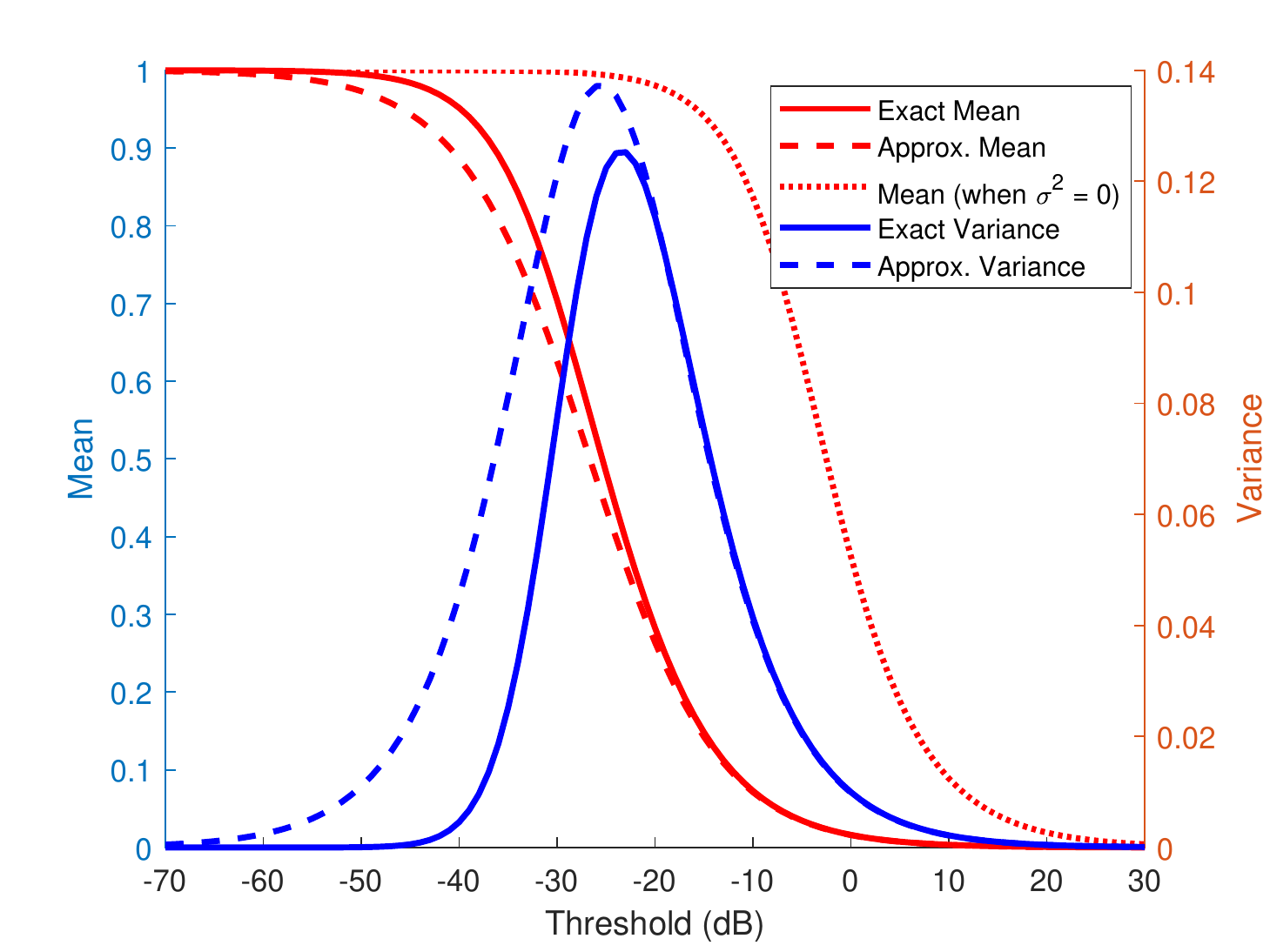}
		\caption{Mean and variance and their approximations for $\gamma = 3$.}
		\label{fig:approx-moments}
	\end{center}
\end{figure}


In Fig. \ref{fig:relative-error}, we plot the relative error between the beta distribution and Fourier-Jacobi expansion. The $\alpha$ and $\beta$ are assigned according to the moment matching method while the Fourier-Jacobi expansion is truncated after ten terms. We observe that Fourier-Jacobi expansion gives more accurate result, and the relative error is under $1\%$ for the most part. 

In Fig. \ref{fig:approx-moments}, we plot the mean and variance of meta distribution and their approximation for $\gamma=3$. As a comparison, the mean for the noiseless system is also given. We see that the approximation has highest error around the knee of the curve.


\section{Conclusion}
We have shown how we can reconstruct the meta distribution given all of its moments via Fourier-Jacobi expansion. We  have also analyzed the error characteristics of a simple approximation for its moments for a Poisson cellular network.


\bibliographystyle{IEEE}

\end{document}